\documentclass[preprint,showpacs,preprintnumbers,amsmath,amssymb]{revtex4}

\usepackage{graphicx}
\usepackage{dcolumn}
\usepackage{bm}

\begin{document}

\preprint{}

\title{Separability Analyses of Two-Qubit Density Matrices}

\author{Paul B. Slater}%
\email{slater@kitp.ucsb.edu}
\affiliation{%
ISBER, University of California, Santa Barbara, CA 93106\\
}%
\date{\today}

\begin{abstract}
We pursue a number of analytical directions, motivated to some extent 
initially by
the possibility of developing a methodology for {\it formally} proving
or disproving a certain conjecture of quantum-theoretical relevance
(quant-ph/0308037).
It asserts that the 15-dimensional volume occupied by the
{\it separable} two-qubit density matrices is $(\sqrt{2}-1)/3$, as measured
in terms of the statistical distinguishability metric (four times the
Bures or minimal monotone metric).
Somewhat disappointingly, however, the several various analyses 
that we report, though we hope 
of independent/autonomous 
interest, appear to provide small indication of how
to definitively resolve the conjecture. Among our studies here 
are ones of: (1) the Bures volumes of
the two-dimensional sections of Bloch vectors for 
a number of the  Jak\'obczyk-Siennicki two-qubit
scenarios; (2) the structure of certain 
convex polytopes of
separable density matrices; and (3) the diagonalization of $15 \times 15$ 
Bures metric tensors.
\end{abstract}

\maketitle

\section{Introduction}
In a previous study \cite{slatersilver}, we formulated --- based on some
combination of numerical and analytic evidence --- a conjecture that the
volume, as measured in terms of the statistical distinguishability (SD)
metric \cite{samcarl}, of the 15-dimensional convex set of 
 $4 \times 4$ {\it separable} density matrices is
\begin{equation} \label{SDconjecture}
V_{SD}^{sep} = (\sqrt{2}-1)/3 \approx 0.138071.
\end{equation}
The numerator of (\ref{SDconjecture}), that is $\sigma_{Ag} \equiv 
\sqrt{2}-1$ (alternatively, $\sqrt{2}+1 = 1/(\sqrt{2}-1)$), has been previously termed the ``silver mean'' \cite{minnesota}.
It is interesting to note that this SD
conjecture (\ref{SDconjecture}) 
involves the first three positive integers alone, while the ``golden mean''
\cite{livio} is defined as $(\sqrt{5}+1)/2$ (or, alternatively as
$(\sqrt{5}-1)/2 = 2/(\sqrt{5}+1)$).

A numerical integration based on two {\it billion} (separable {\it and} 
nonseparable) density matrices generated by 
 a quasi-Monte Carlo (Faure-Tezuka) procedure 
had yielded an estimate of 0.137884 for $V^{sep}_{SD}$ 
\cite{slatersilver}. (It now appears \cite{qubitqutrit} that it is possible
to significantly accelerate the MATHEMATICA program employed, so that we may
soon be able to report results for substantially larger sample sizes.) In
 \cite{slaterC}, an exact  Bures {\it probability} of separability 
($V^{sep}_{Bures}/V^{sep+nonsep}_{Bures}$) equal to
$\sigma_{Ag}$ had been obtained
 for both the $q=1$ and $q = \frac{1}{2}$ 
two-qubit states \cite{abe}
inferred using the principle of maximum nonadditive [Tsallis]
entropy --- and also for an additional {\it low}-dimensional 
scenario \cite[sec. 2.2.1]{slaterC}.

Since the {\it Bures} (minimal monotone \cite{petzsudar}) 
metric \cite{hubner1,hubner2} 
is identically one-fourth the SD
metric \cite{samcarl}, that is
\begin{equation}
 d_{Bures}(\rho,\rho+ d \rho)^2 =
(1/4) d_{SD}(\rho,\rho +d \rho)^2,
\end{equation}
in the neighborhood of a density matrix $\rho$, 
the conjecture (\ref{SDconjecture}) 
becomes equivalent to one that the Bures volume
is 
\begin{equation}  \label{silvermean}
V_{Bures}^{sep} = 2^{-15} (\sqrt{2}-1)/3 \approx 4.2136 \cdot 10^{-6}.
\end{equation}
(In \cite{slatersilver}, we had also been led to a number of related
conjectures, including one that $V^{sep}_{KM} =30 V^{sep}_{Bures}$, for the 
two-qubit systems, 
where $KM$ denotes the Kubo-Mori monotone metric.)

In the present
 analysis, we undertake a line of research that hopefully --- we, at least,
initially thought --- may
contribute to 
 {\it formally} proving/disproving this conjecture. This has
seemed a quite formidable task, as a 
(naive ``brute force'') 
{\it symbolic} integration approach,
along the lines of the 
high-dimensional {\it numerical} integration followed in \cite{slatersilver},
appears to be far beyond present computing technology (cf. \cite{qubitqutrit}).
In this previous work \cite{slatersilver}, we generated $4 \times 4$ density
matrices, which then had to be checked for possible separability. Here, our
approach focuses at the outset on separable density matrices, 
and the nonseparable
ones do not directly enter the picture (cf. \cite{aravind,ericsson,kus}).

The convex domain ($D$) of $4 \times 4$ separable density matrices is
15-dimensional in nature.  In sec.~\ref{sectA} we 
systematically generate 16 pure product
$4 \times 4$ separable density matrices,  convex
combinations of  these sixteen spanning some subset of $D$. 
(The entire domain $D$ itself, however, 
is the closed convex hull of the set of {\it all} --- not just 16 --- product
states \cite{clifton}.)
This approach can be 
interestingly contrasted (in sec.~\ref{sectD}) 
with that of Braunstein {\it et al}
\cite{samNMR}, who used --- in the $4 \times 4$ case --- an
{\it overcomplete} basis of 36 density matrices to ``give a constructive 
proof that all mixed states of $N$ qubits in a sufficiently small
neighborhood of the maximally mixed state are separable''. They allowed
{\it negative} weights on their basis matrices in their analysis, 
thus enabling them to obtain {\it entangled} density matrices.
We view the convex weights ($w_{i}, i = 1,\ldots,15$) attached to the first
of our 
15 matrices as the parameters or variables of our problem. (Of course, then,
we must have $w_{16}=1 -\Sigma_{i=1}^{15} w_{i}$ as a bound or dependent
parameter.) 

Proceeding onward, in
sec.~\ref{sectB} we investigate the possibility of using certain of the
results in sec.~\ref{sectA} to obtain bounds on $V^{sep}_{Bures}$.
We examine in sec.~\ref{BMT} the associated $15 \times 15$ Bures metric tensor, and in sec.~\ref{sectD} consider the use of an overcomplete basis.
In sec.~\ref{sectE}, we generate a polytope composed of $8 \times 8$ 
{\it three}-qubit density matrices, while we obtain in sec.~\ref{Diagonal}
the Bures metric tensor in diagonal form. We obtain and plot 
in sec.~\ref{sectG} 
the Bures volume elements for three of the two-dimensional two-qubit
scenarios presented by Jak\'obczyk and Siennicki \cite{jakobczyk}, and 
derive the associated  ``Euclidean'' probabilities of separability.
\section{Analyses}
\subsection{Convex Polytope of 
$4 \times 4$ Density Matrices} \label{sectA}
To generate the 16 
extreme pure product 
basis matrices (cf. \cite{DMSST,pittenger}), 
we start with four $2 \times 2$ density matrices,
the Bloch vectors of each of which extend to one of the four points 
$((v,v,v), (-v,-v,v), (-v,v,-v)$ and $(v,-v,-v)$, where $v=1/\sqrt{3}$) 
of
an inscribed tetrahedron (a 
Platonic solid). Then, we obtain 16 extreme points of $D$, by taking
all possible tensor products of pairs of 
these four $2 \times 2$ density matrices.
(We note that Schack and Caves \cite[sec. 2.4]{schackcaves}
mentioned the possibility of this construction, among others, 
observing that
the ``4 projectors for a tetrahedron are linearly independent, making
the corresponding tetrahedral representation [of a $2 \times 2$ density
matrix] unique''.)

\begin{figure}
\includegraphics{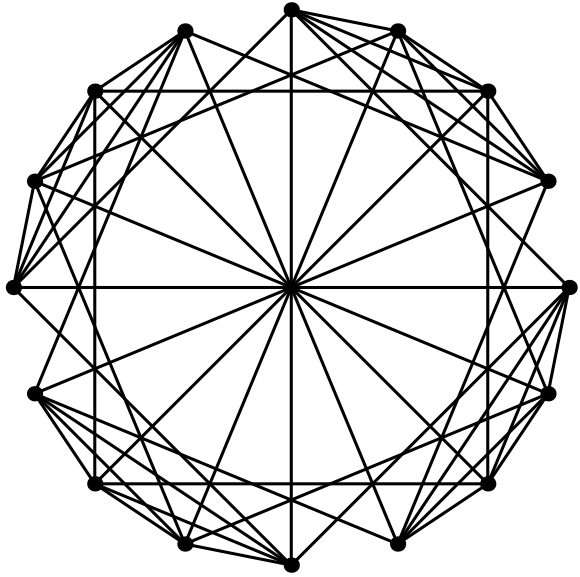}
\caption{\label{fig:graph1} Graph representation of the 48 links between
pairs of 
the 16 extreme pure product separable density matrices 
corresponding to a Bures
distance equal to $\sqrt{2 -2/\sqrt{3}} \approx 0.919402$ (and Hilbert-Schmidt
distance of $2/\sqrt{3} \approx 1.1547$). The 
16 nodes are grouped into four quartets.}
\end{figure}

These 16 extreme pure product density matrices have an interesting structure.
Each one is a Bures distance of $\sqrt{2-2/\sqrt{3}} \approx 0.919402$
(and a Hilbert-Schmidt distance 
\cite{karolhans} of $2/\sqrt{3} \approx 1.1547$) 
from {\it six} other matrices, 
and a Bures distance of $2/\sqrt{3} \approx 1.1547 $ (and a HS distance of
$4/3 \approx 1.3333$) from the {\it nine} remaining matrices.
(Using the trace of the matrix product of pairs of these matrices,
as an alternative distance measure, we
get the results 1/9 [for six] and 1/3 [for nine].)
The Bures {\it 
distance} between two density matrices ($\rho_{1}$ and $\rho_{2}$),
given by the formula,
\begin{equation}
d_{Bures}(\rho_{1},\rho_{2}) = \sqrt{2 -2 \mbox{Tr}[ (\rho_{1}^{1/2} \rho_{2} 
\rho_{1}^{1/2})^{1/2}}],
\end{equation}
can only yield distances  in the range $[0,\sqrt{2}]$.
The Hilbert-Schmidt distance is defined as  \cite[eq. (2.3)]{karolhans}
\begin{equation}
d_{HS}(\rho_{1},\rho_{2}) = ||\rho_{1} -\rho_{2}||_{HS} = \sqrt{\mbox{Tr}[(\rho_{1}-\rho_{2})^2]}.
\end{equation}
Let us  also observe that an explicit distance 
formula is presently available for just 
one other
``monotone'' metric than the Bures, 
that is, the Wigner-Yanase metric \cite[eq. (5.1)]{pablo},
namely,
\begin{equation}
d_{WY}(\rho_{1},\rho_{2})^2 = 
4 \mbox{arccos}^2 {\Big( \mbox{Tr}[ 
\rho_{1}^{1/2} \rho_{2}^{1/2}] \Big)}.
\end{equation}
(The class of all monotone metrics has, of course, a nondenumerable
number of members.)
In Fig.~\ref{fig:graph1},  we show a graph representation --- in which only  
the
links 
corresponding to the Bures distance $\sqrt{2-2/\sqrt{3}} \approx 0.919402$ are
allowed  --- with 
the 16 extreme density matrices
serving as the nodes. (The {\it spectrum} \cite{cvetkovic,biggs} 
 of the associated (0,1)-adjacency
matrix consists of the integers 
6, -2 [nine-fold] and 2 [six-fold].)

\begin{figure}
\includegraphics{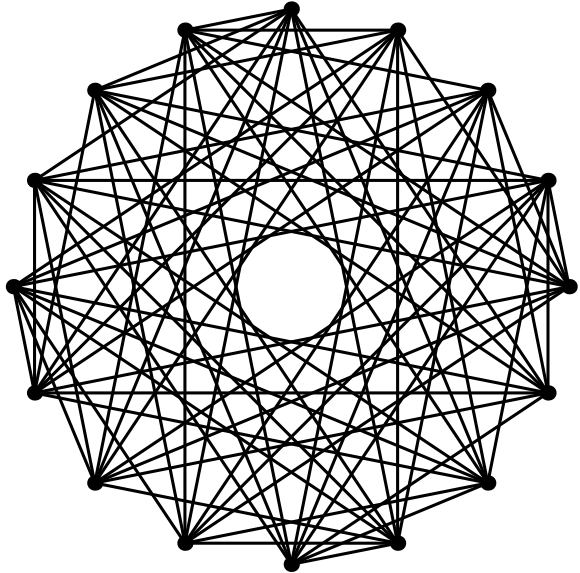}
\caption{\label{fig:graph2}  Graph representation of the 72 links between
pairs of 
the 16 extreme pure product separable density matrices corresponding
to a Bures
distance equal to $2/\sqrt{3} \approx 1.1547$ (and Hilbert-Schmidt
distance of $4/3 \approx 1.33333$.)}
\end{figure}

In Fig.~\ref{fig:graph2},  we show a graph representation --- in which, 
now,  only those  edges corresponding to the Bures distance of
 $2 / \sqrt{3}  \approx 1.1547$ are
displayed. (Its spectrum 
consists of the integers 9, 1 [nine-fold] and -3 [six-fold].)

\subsection{Volume calculations} \label{sectB}

Additionally, the 16 extreme density matrices (and, it would seem any 
pure state, in general) are each a Bures distance of
1 (and a HS distance of $\sqrt{3}/2 \approx 0.866025$)
 from the fully mixed state $I_{4}$. 
(The WY-distance from each of the extreme 16 matrices to $I_{4}$ is 
$2 \pi/3$.) 
Now, a 15-dimensional
{\it Euclidean} ball of radius 1 has a volume equal to 
$256 \pi^7 /2027025 \approx 0.381443$  and one of radius
$\sqrt{3}/2 \approx 0.866025$, a volume equal to $27 \sqrt{3} \pi^7 /3203200 
\approx 0.044095$. These Euclidean 
volumes are certainly
{\it larger} than the Bures and
HS volumes 
of the separable {\it plus} nonseparable $4 \times 4$
density matrices. These two latter values (precisely known only recently,
making use of the theory of random matrices 
 \cite[eq. (4.12)]{hanskarol} \cite[eq. (4.5)]{karolhans})
are,  respectively,
\begin{equation}
V_{Bures}^{sep+nonsep} = \frac{\pi^{8}}{165150720}
\approx 0.0000574538; \hspace{.5cm} 
V_{HS}^{sep+nonsep} = \frac{\pi^{6}}{851350500} \approx 1.12925  \cdot 10^{-6}.
\end{equation}
(Also, in \cite{slatersilver}, we had been led to conjecture that, in the
two-qubit case, $V^{sep+nonsep}_{KM} =64 V^{sep+nonsep}_{KM}$, where 
 denotes the Kubo-Mori monotone metric.) 
This Bures volume, remarkably, is the same as 
the 15-dimensional volume (``hypersurface area'')
of a 15-dimensional Euclidean {\it hemisphere} (cf. \cite{uhlmann})
with radius 1 \cite{hanskarol,slatersilver}.
(In the numerical integration \cite[sec.~1.2]{slatersilver} based on two billion sampled
points, a very close 
estimate of 0.000574536 was obtained for $V^{sep}_{Bures}$.)
Coupling this with the ``silver mean'' conjecture (\ref{silvermean}) 
easily yields an
implied conjecture that the Bures ``probability of separability'' 
\cite{ZHSL} is
\begin{equation}
P^{sep}_{Bures} = \frac{V_{Bures}^{sep}}{V_{Bures}^{sep+nonsep}}=
\frac{(1680 (\sqrt{2}-1)}{\pi^8} \approx 0.0733389.
\end{equation}
The {\it scalar curvature}  of the Bures metric 
at the fully mixed state $I_{4}$ is known --- as a specific case of more
general formulas \cite[Cor. 3]{dittscalar} --- to equal 570.
We can, thus,  attempt to 
apply  the formula \cite[p. 3676]{andai} (cf. \cite[p. 936]{petz}),
\begin{equation} \label{AndaiSC}
V(B_{r}(\rho_{0}))= \frac{\sqrt{\pi^{n^2-1}} r^{n^2-1}}{\Gamma{(\frac{n^2+1}{2})}}
\cdot \Big(  1- \frac{\zeta (\rho_{0}) r^2}{6 (n^2+1)} +O(r^4) \Big) ,
\end{equation}
for the volume of a geodesic ball centered at the 
$n \times n$ density matrix $\rho_{0}$,
where $\zeta (\rho_{0})$ is the scalar curvature at $\rho_{0}$.
(The scalar curvature of the [monotone] Wigner-Yanase metric is a 
{\it constant} over the domain of $n \times n$ density matrices
equal to $\frac{1}{4} (n^2-1) (n^2-2)$ 
\cite[Thm. 7.2]{pablopreprint}, which gives us ``only'' $52.5$ for $n=4$.)
Taking the radius of the ball as 
$r=1$, and $\zeta (\rho_{0}) =570$, $n=4$ and $\rho_{0}=I$,
we get the result $V(B_{1}(I_{4})) = 
512 \pi^7/883575 \approx - 1.75015$ plus a higher-order
term. (Note the {\it negative} sign!
So, it appears that more explicit 
terms in the asymptotic expansion (\ref{AndaiSC})
are needed to try to avoid this. In his dissertation, A. Andai 
has such an expansion (with an $O(r^8)$ term),
but considerably more differential-geometric information is needed beyond
the scalar curvature to implement it (cf. \cite{gray}).

If we fully mix, not the 16 extreme matrices, but only 15 of them in turn
(the 15 nonzero $w_{i}$'s all equalling 1/15),
we find that the resulting 16 
density matrices are all a Bures distance of 
$\sqrt{2 -2 \sqrt{3/5} -1/\sqrt{5}} \approx 0.0599422$ (and a HS-distance 
of $1/(10 \sqrt{3}) \approx 0.057735$)  from 
 $I_{4}$. A 15-dimensional Euclidean ball with this radius has
a volume of only 
\newline
$ 256 (10 -\sqrt{5} -2 \sqrt{15})^{15/2} 
\pi^7 /(158361328125 \sqrt{5}) \approx 1.76774 \cdot 10^{-19}$
(the comparable HS figure being $1.00698 \cdot 10^{-19}$).
Corresponding
 application of the formula (\ref{AndaiSC}) gives us a quite similar
estimate of the Bures volume 
of $1.73225 \cdot 10^{-19}$ (plus a higher-order term) of our 15-dimensional
convex body of separable states.

If we fully mix only 14 of the 16 (the 14 nonzero $w_{i}$'s all
equalling 1/14), then the resultant Bures distance from
$I_{4}$ is {\it either} 0.0810507 (for 72 of the possible pairs) or 0.0973228
(for the other 48 possible pairs). (The comparable HS-distances are
$\sqrt{11}/42 \approx 0.0789673$ and $(\sqrt{5/3})/14 \approx 0.0922139$.)
\subsection{Bures metric tensor} \label{BMT}
Using our  convex geometry/parameterization, 
we can further compute \cite[Prop. 2]{dittcompute} the $15 \times
15$ Bures metric tensor ($||g_{ij}||$) 
at $I_{4}$.
(This fully mixed state is obtained by setting $w_{i} =1/16$ for all $i$,
 in forming the convex combination of the 16 extreme 
product basis matrices.)
It has {\it five} eigenvalues equal 
to $17/s$,  where $s =98304 = 2^{15} \cdot 3$, 
{\it eight}
 equal to $17/(3 s)$ and
a {\it pair} $17 (31 \pm \sqrt{769})/(6 s)$, 
having the 
approximate 
values 0.00169725 and 0.000094224, being the roots of the
quadratic equation
\begin{equation} \label{quadratic}
(2^{26} \cdot 3^3) x^2 -(2^{11} \cdot 3 \cdot 7 \cdot 31) x
+17^2 =0.
\end{equation}
If we transform this tensor by the Jacobian 
(the determinant of which equals $2^{14} \sqrt{2}$) corresponding to returning 
from our choice of
basis variables to that relying  upon the naive parameterization 
of a two-qubit density matrix, say,
\begin{equation} \label{naive}
\begin{pmatrix} 
a_{11}  & a_{12} + b_{12} i & a_{13} + b_{13} i & a_{14} + b_{14} i \\
a_{12} - b_{12} i & a_{22} &  a_{23} + b_{23} i & a_{24} + b_{24} i \\
a_{13} - b_{13} i & a_{23} - b_{23} i & a_{33} & a_{34} + b_{34} i \\
a_{14} - b_{14} i & a_{24} - b_{24} i & a_{34} - b_{34} i & 1- a_{11} -a_{22} -a_{33},
\end{pmatrix}
\end{equation}
we can obtain the invariant 
trace of the Bures metric tensor (cf. \cite{frieden}), which is $255/2^{16}$.
The square root of the determinant of the transformed tensor 
gives us the 
invariant {\it volume element}
at $I_{4}$. This we computed to equal $2^{-120} 17^7 \sqrt{17/2}
\approx 9.00021 \cdot 10^{-28}$.

\begin{figure}
\includegraphics{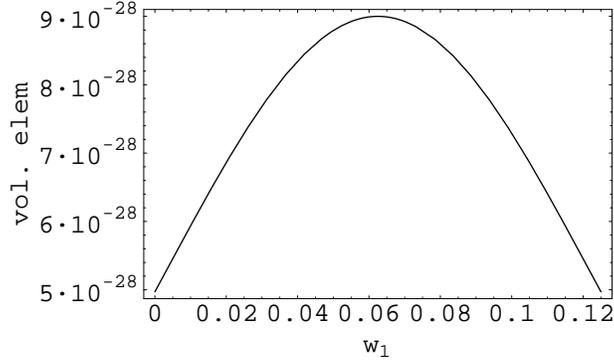}
\caption{\label{fig:volelem} 
Plot of volume element of $15 \times 15$ Bures metric tensor
in the neighborhood of the fully mixed state $I_{4}$,
as a function of $w_{1}$ over the range [0,1/8], 
with $w_{i}=1/16$ for $i=2,\ldots,15$ and
$w_{16}=1/8-w_{1}$. For $w_{1}=w_{16}=1/16$, one gets $I_{4}$.
The function peaks at $I_{4}$, having the exact value $2^{-120} 
17^7 \sqrt{17/2}
\approx 9.00021 \cdot 10^{-28}$
there.}
\end{figure}

\begin{figure}
\includegraphics{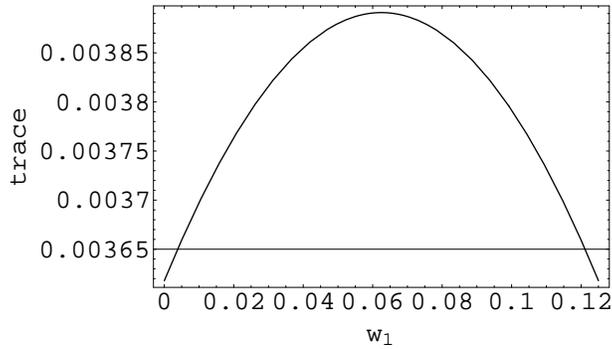}
\caption{\label{fig:trace} Plot of trace
 of $15 \times 15$ Bures metric tensor
in the neighborhood of the fully mixed state $I_{4}$,
as a function of $w_{1}$ over the range [0,1/8],
with $w_{i}=1/16$ for $i=2,\ldots,15$ and
$w_{16}=1/8-w_{1}$. For $w_{1}=w_{16}=1/16$, one gets
 $I_{4}$. The function peaks at $w_{1}=1/16$ with the value
$255/2^{16} \approx 0.00389099$ there.}
\end{figure}

If we allow the weight ($w_{1}$) assigned to the first, say, of the 16
matrices to vary between 0 and 1/8, and the weight ($w_{16} =1/8-w_{1}$) 
compensatingly likewise, keeping the fourteen 
other weights as 1/16, 
the determinant of the $15 \times 15$ 
Bures metric tensor is a (high) 80-degree
polynomial in $w_{1}$, but the graph (Fig.~\ref{fig:volelem}) 
of its square root multiplied by the Jacobian determinant (that is, 
the corresponding
volume element $|g_{ij}|^{1/2}$) seems
 quite well-behaved --- clearly  peaking at $w_{1}=w_{16}=1/16$, that is at
$I_{4}$. The integral of this function over
[0,1/8] is $9.38545  \cdot 10^{-29}$ --- a result certainly 
 not inconsistent with
our ``silver mean'' conjecture (\ref{silvermean}) for the {\it total}
Bures volume $V_{Bures}^{sep}$ 
of $D$, the full 15-dimensional convex set of $4 \times 4$
separable states.
The Bures distance between the two extremes on this plot 
(that is, one state with 
$w_{1}=0, w_{16} =1/8$ and one with 
$w_{1}=1/8, w_{16}=0$, while all other
$w$'s remain fixed  at 1/16) is $(6-\sqrt{34})/6 \approx 0.0281747$, 
considerably larger than the indicated integral between these two endpoints.
In Fig.~\ref{fig:trace}
 we show the same form of plot 
as Fig.~\ref{fig:volelem}, but now 
using the {\it trace} of the $15 \times
15$ metric tensor rather than the volume element.

We were not able to compute determinants of the $15 \times 15$ metric
tensor when {\it more}
 than one of the convex weights was allowed to freely vary.
(We were also unable to express a Bell state as a linear combination --- of
course, requiring negative weights --- of
the 16 extreme pure product density matrices.)

\subsection{Overcomplete Basis} \label{sectD}
\begin{figure}
\includegraphics{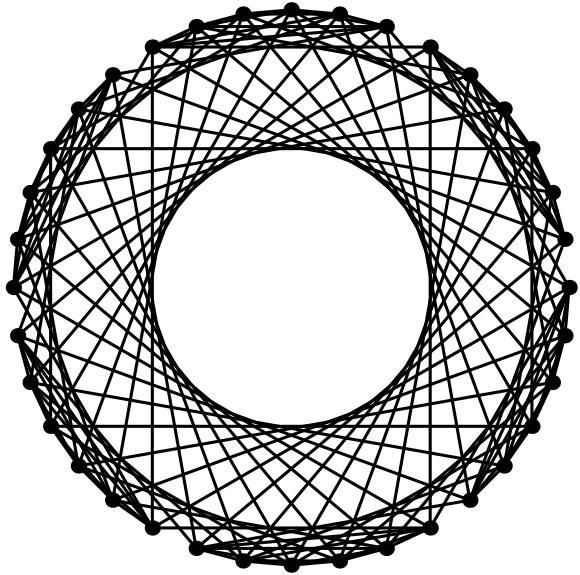}
\caption{\label{fig:graph4} Graph representation of the
144 links between pairs of
the 36 extreme pure product separable density matrices --- based on the
{\it overcomplete} ``Pauli'' basis of Braunstein {\it et al} 
\cite{samNMR} for the two-qubit case --- corresponding to a Bures
distance equal to $\sqrt{2 -\sqrt{2}}$. The nodes are grouped into sextets.}
\end{figure}

If we were to use in the two-qubit case, not the 16 extreme density matrices
employed above, but the 36 extreme density matrices (forming an overcomplete
basis), applied by Braunstein {\it et al} \cite{samNMR}, then we find
that of the $35 \cdot 36 /2 =630$ possible distinct unordered pairs of
density matrices, 144 correspond to a Bures distance of $\sqrt{2-\sqrt{2}}$,
198 to $\sqrt{2}$ and 288 to 1. (Thus, as opposed to the analysis based on the
16 matrices, there are {\it three}, not {\it two} different 
sets of internodal distances.)
In Fig.~\ref{fig:graph4}, we represent those 144 links with a Bures distance
of $\sqrt{2-\sqrt{2}}$.
\subsection{Convex Polytope of $8 \times 8$ Density Matrices} \label{sectE}
We also extended the  general 
approach above to the {\it three}-qubit 
case, by taking tensor
products of all possible {\it triplets} of the 
same four $2 \times 2$ density
matrices with 
their 
Bloch vectors reaching the points of the inscribed tetrahedron.
(The extension to the qubit-{\it qutrit} case seems more problematical,
since there seems to be no natural immediate analogue of the tetrahedron to
employ for choosing {\it nine} 
 $3 \times 3$ density matrices to tensor product with
the {\it four} $2 \times 2$ density matrices we have been utilizing.)
This gives us 64 extreme product {\it triseparable} states.
There are, then, $63 \cdot 64/2 = 2016$ distinct pairs of intermatrix 
Bures distances.
Of these, 288  are $\sqrt{2 -2/\sqrt{3}} \approx 0.919402$, 864 are
$2/\sqrt{3} \approx 1.1547$ and another 864  are $\sqrt{2 -2 /(3 \sqrt{3})}
\approx 1.27087$.
(The corresponding HS-distances are $2/\sqrt{3} \approx 1.1547$, $4/3 
\approx 1.33333$ and $2 \sqrt{13/3}/3 \approx 1.38778$.)

\begin{figure}
\includegraphics{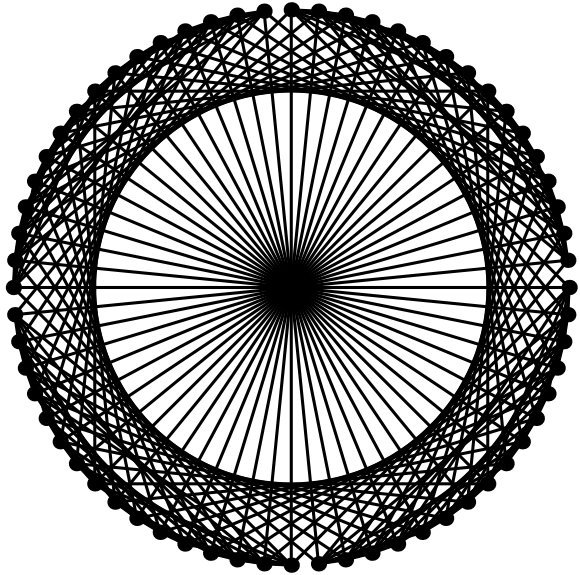}
\caption{\label{fig:graph3}  Graph representation of the 
288 links between pairs of
the 64 extreme pure product triseparable density matrices having Bures
distance equal to $\sqrt{2 -2/\sqrt{3}} \approx 0.919402$ (and Hilbert-Schmidt
distance equal to $2/\sqrt{3} \approx 1.1547$). The 
64 nodes fall into 4 groups
of 16.}
\end{figure}

In Fig.~\ref{fig:graph3},  
we show a graph representation --- in which only the 
edges corresponding to the Bures distance 
$\sqrt{2-2/\sqrt{3}} \approx 0.919402$ are
allowed  --- with
the 64 extreme density matrices
serving as nodes. (The spectrum of the associated (0,1)-adjacency
matrix consists of 
the integers 
9, 5 [nine-fold], -3 [27-fold] and 1 [27-fold]. For the 
graph based on the Bures distance $2/\sqrt{3}$, the spectrum has 27, -5 [27-fold], and 3 [36-fold] and that based on $\sqrt{2-2/(3 \sqrt{3})}$, 27, 
-9 [nine-fold], 3 [27-fold] and -1 [27-fold].)
Graph 
representations 
with the other two possible links are perhaps too dense visually
for a meaningful representation.

The 64 extreme matrices are all a Bures distance of $\sqrt{2-1/\sqrt{2}}
\approx 1.13705$ (and a HS-distance of $\sqrt{7/2}/2 \approx 0.935414$) 
from the fully mixed state $I_{8}$. 
{\it Euclidean} 35-dimensional balls of these radii
have volumes of 0.0000299764 and $3.23277 \cdot 10^{-8}$, respectively.
(The formula (\ref{AndaiSC}) --- taking into account that the scalar
curvature at $I_{8}$ 
 equals 9954 \cite{dittscalar}--- gives us 
a ``volume'' $1.03199 \cdot 10^{-14}$, but
with a {\it minus} sign.)
The trace of the associated $63 \times 63$ Bures metric tensor was
found at $I_{8}$ to equal $7/2^{74}$.
The square root of the determinant of the tensor 
equals $(2^{2359} 3^{72} \sqrt{2})^{-1}$.
(Multiplication by the determinant of the Jacobian of the transformation back
to the naive parameterization of the $8 \times 8$ density matrices (cf. (\ref{naive}))
would give us the volume
element at $I_{8}$.)
The spectrum of the tensor consists of 26 eigenvalues equal to
$1/(2^{75} 3^3)$, 26 equal to $1/(2^{75} 3^2)$ and 8 equal to $(1/(2^{75} 3)$ --- plus three ($1.90525 \cdot 10^{-22}, 
 1.3992 \cdot 10^{-24}$ and $6.10756 \cdot 10^{-24}$) --- that 
are the roots of the cubic equation,
\begin{equation} 
(2^{219} \cdot 3^6) x^3 - (2^{145} \cdot 3^3  \cdot 101) x^2
+(2^{69} \cdot 3 \cdot 499) x -1 =0.
\end{equation}
We have not, however, yet been able to compute the associated eigenvectors.
\subsection{Diagonal $15 \times 15$ Bures Metric Tensor} \label{Diagonal}
We were not only able --- using MATHEMATICA --- to find the (four distinct)
eigenvalues, given above in sec.~\ref{BMT}, 
using the parameterization based on
 the 16 extreme pure product 
density matrices, of the $15 \times 15$ Bures metric tensor at
$I_{4}$, but also the associated set of orthonormal eigenvectors.
Using the components of these eigenvectors as weights, we performed a 
linear 
transformation from the variables $w_{i}$ to a new set, call them $v_{i}$, so
that the metric tensor was now {\it diagonal} (cf. \cite{tod,cao}). Additionally, at $I_{4}$, only
two of the $v_{i}$'s were nonzero --- let us denote them $v_{1}$ 
and $v_{7}$ (as this is how they arise in the analysis) --- and 
these were precisely the ones associated
with the paired eigenvalues that are the roots of the quadratic equation
(\ref{quadratic}).
So, at $I_{4}$, all $v_{i}$'s are zero, but for two that are the 
square roots
of the roots of the quadratic equation
\begin{equation} \label{quadratic2}
(2^9 \cdot 769) x^2 - (2 \cdot 3 \cdot 5 \cdot 769) x - 3^3=0.
\end{equation}
These square  roots are $\sqrt{\frac{15}{512} \pm \frac{399}{512 \sqrt{769}}}$, having the approximate values of 0.239581 and 0.0345646.
(The larger value 
is associated with the eigenvector for the 
larger root of (\ref{quadratic}).)
The 15 $4 \times 4$ density matrices weighted by the 15 $v_{i}$'s 
in the 
{\it new }
expansion of a $4 \times 4$ density matrix were all 
{\it traceless} $4 \times 4$ 
Hermitian matrices. (They were mutually orthogonal, but their self-inner products did not all have the same value, though certainly they could be rescaled
so that they do. If we do, in fact, rescale so that all self-inner products
equal 2, then we have that $v_{1},v_{7}=\frac{1}{4}
\sqrt{3 \pm \frac{83}{\sqrt{769}}}$.) There was also a
``constant'' term (weighted by no parameter $v_{i}$) 
in this expansion, which was the density matrix of a pure state.
We can write this density matrix 
as $|j \rangle \langle j|$, where
\begin{equation}
|j \rangle = \frac{1}{\sqrt{6 (2 +\sqrt{3})}} (-\frac{i (5 + 3 \sqrt{3})}{1+\sqrt{3}},
(-\frac{1}{2} +\frac{i}{2})(1+\sqrt{3}),(-\frac{1}{2} +\frac{i}{2})(1+\sqrt{3}),1).
\end{equation}
Just as we completed the analysis immediately above, we noticed the
appearance of a preprint of Kimura and Kossakowski \cite{kk}
(cf. \cite{byrdkhaneja})
in which the (somewhat similar) 
density operator expansion
\begin{equation} \label{LieExpansion}
\rho= \frac{\mbox{Tr} \rho}{N} I_{N} +\frac{1}{2} \Sigma_{i=1}^{N^2-1}
(\mbox{Tr} \rho \lambda_{i}) \lambda_{i}
\end{equation}
is proposed.
Here the $\lambda$'s ($i=1,\ldots,N^2-1$) 
are orthogonal generators of $SU(N)$
which satisfy
\begin{equation}
\mbox{(i)} \lambda_{i}^{\dagger} = \lambda_{i}, \mbox{(ii)}
\mbox{Tr} [\lambda_{i}]=0,
(\mbox{iii}) \mbox{Tr} [\lambda_{i} \lambda_{j}]= 2 \delta_{ij}.
\end{equation}
We, then, reran our computer programs using this parameterization (employing
the standard generators of $SU(4)$). We found that
for the fully mixed state  $I_{4}$ (where all 
$\langle \lambda_{i} \rangle = \mbox{Tr}[\rho \lambda_{i}]$ 
are equal to zero),
 not only that the $15 \times 15$ Bures metric tensor
was {\it diagonal}, but also that all its 
 diagonal entries (eigenvalues) were equal to a {\it common} 
value, that is, 17/262144, 
where $262144 =2^{18}$. The square root of the determinant 
of the tensor was, then,  equal to $(17/262144)^{15/2}$, which upon 
multiplication by the determinant $2^{14} \sqrt{2}$ of the Jacobian
for the change-of-variables back to the naive parameterization 
(\ref{naive}) of the 
$4 \times 4$ density matrices, gives us again the invariant volume
element $2^{-120} 17^7 \sqrt{17/2}
\approx 9.00021 \cdot 10^{-28}$.

Also, when we let one parameter vary freely ($\langle \lambda_{1} \rangle $, 
say) 
from 0, the Bures metric tensor remained diagonal, unlike the situation in
our earlier analysis.
However, it appeared that if we were to let all 15 of the parameters vary
freely, then the $15 \times 15$ Bures metric tensor is {\it not}
diagonal. (We generated the [very large] symbolic 
entries of this tensor and then
substituted random values of the 15 parameters, in reaching this conclusion.)
This has the practical, as well as theoretical significance that one can
readily compute the determinant of a symbolic multivariate diagonal
$15 \times 15$ matrix, but certainly not one with many off-diagonal
nonzero entries.

The $15 \times 15$ Bures metric tensor obtained using the naive parametrization
(\ref{naive}) is itself quasi-diagonal, with the only off-diagonal structure
being the $3 \times 3$ submatrix pertaining to the diagonal entries
$a_{11},a_{22},a_{33}$. The six off-diagonal nonzero entries are $2^{-17}$
and the 15 diagonal entries are $2^{-16}$. The volume element (square root
of the determinant) at 
$I_{4}$ is, as mentioned previously, $2^{-120} 17^7 \sqrt{17/2}$.

\subsection{Jak\'obczyk-Siennicki Two-Qubit Analyses} \label{sectG}
Jak\'obczyk and Siennicki \cite{jakobczyk}
studied --- using the same $SU(4)$ 
parameterization (\ref{LieExpansion}) as employed by Kimura 
and Kossakowski \cite{kk} (which we used to obtain 
a {\it diagonal equal-entry}
$15 \times 15$ Bures metric tensor 
[sec.~\ref{Diagonal}]) --- {\it two}-dimensional 
sections of a set of Bloch
vectors corresponding to two-qubit density matrices.
In an interesting set of diagrams, they depicted
both the domains of these  $4 \times 4$ density matrices
\cite[Fig. 1]{jakobczyk} and their convex subsets of separable states 
\cite[Fig. 2]{jakobczyk} (in those specific 
cases in which
there were nontrivial inclusions). 

\begin{figure}
\includegraphics{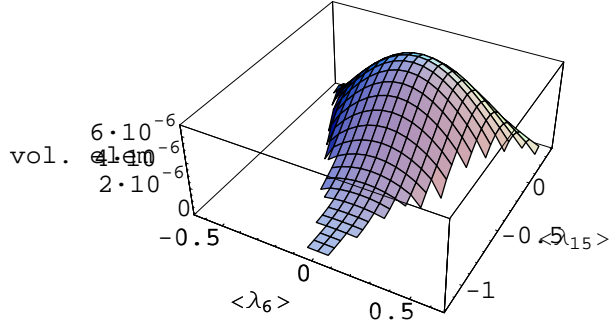}
\caption{\label{fig:graphJakobC} Bures volume element over the 
{\it triangular} domain of
two-parameter $4 \times 4$ (separable and nonseparable) 
density matrices, denoted C by
Jak\'obczyk and Sennini \cite[Fig. 1]{jakobczyk}. The Bures volume was
computed to be $6.43885 \cdot 10^{-7}$. The fully mixed state $I_{4}$ is at
the origin.}
\end{figure}

We have, first,  analyzed the case they denote by the letter C, 
in which only the {\it sixth}
and {\it fifteenth} $SU(4)$ generators are assigned nonzero weight in the 
density matrix expansion (\ref{LieExpansion}).
We found that the total Bures volume (of the indicated triangular
domain) was
$3.27995 \cdot 10^{-6}$. Then, using further results of theirs, we computed
the Bures volume of the separable subset  
to be  $6.43885 \cdot 10^{-7}$. 
Thus, the probability of separability (the ratio of these two quantities)
\cite{ZHSL,slaterEuro} for this particular scenario (labeled 
CK') was
0.196309. (Viewing the 
corresponding illustration of the separable and nonseparable 
regions  in \cite[Fig. 2]{jakobczyk} simply naively 
as a Euclidean diagram,
we obtain a ``Euclidean probability of separability'' that is much
higher, being the ratio of $1/\sqrt{6} + \sqrt{2} \pi/9$ to
$2^{5/2}/3^{3/2}$, that is,
$(9 + 2 \pi \sqrt{3})/24 \approx 0.82845$. Also, we were not able to compute
enough digits of accuracy to meaningfully address the question of whether 
these quantities might possibly correspond to analytically exact
expressions.)
In Fig.~\ref{fig:graphJakobC} 
we show the Bures volume element over this 
two-dimensional triangular domain of 
$4 \times 4$ density matrices. 
(The fully mixed state $I_{4}$ corresponds to the origin.)

\begin{figure}
\includegraphics{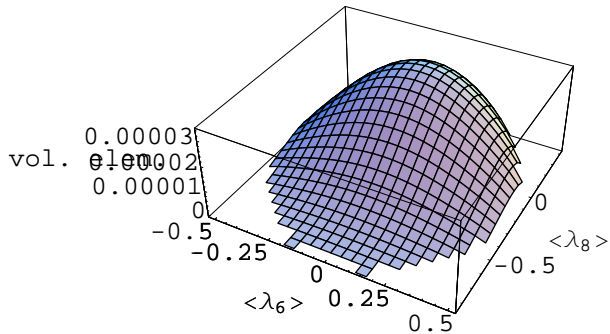}
\caption{\label{fig:graphJakobG} Bures volume element over the
{\it elliptical} domain of
two-parameter $4 \times 4$ (separable and nonseparable) 
density matrices, denoted G by
Jak\'obczyk and Sennini \cite[Fig. 1]{jakobczyk}.}
\end{figure}

In Fig.~\ref{fig:graphJakobG} we analogously show the Bures volume element for 
the two-dimensional scenario Jak\'obczyk and Siennicki denote by the letter
G (in which only the {\it sixth} and {\it eighth} $SU(4)$ generators 
are allowed to bear nonzero
weight).
As a Euclidean figure, the area of this elliptical 
domain is $(9 \pi \sqrt{3/2}) /32 \approx 1.08215$, 
while the area of the separable subdomain is
$(9 (2+\pi) \sqrt{3/2})/64 \approx 0.885535$. 
The naive Euclidean probability of separability is, then, 
simply $(2 + \pi)/(2 \pi) \approx 0.81831$.
Our attempts to compute with reasonable confidence 
the corresponding Bures volumes 
(and, thus, the Bures probability of separability) were impeded by
considerable MATHEMATICA numerical integration difficulties we have not yet
resolved.

\begin{figure}
\includegraphics{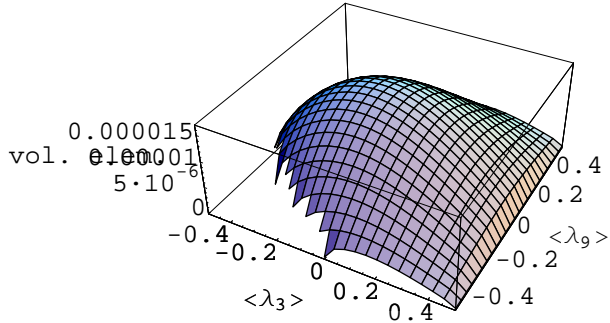}
\caption{\label{fig:graphJakobE} Bures volume element over the
{\it hyperbolic} domain of
two-parameter $4 \times 4$ (separable and nonseparable) 
density matrices, denoted E by
Jak\'obczyk and Sennini \cite[Fig. 1]{jakobczyk}.}
\end{figure}

In Fig.~\ref{fig:graphJakobE} we further show the Bures volume element for
the two-dimensional scenario Jak\'obczyk and Siennicki denote by the letter
E (in which only the {\it third} and {\it nineth} $SU(4)$ generators
are allowed to bear nonzero
weight).
As a Euclidean figure, the area of this hyperbolic
domain is $2 \sqrt{2}/3 \approx 0.942809$,
while the area of the separable subdomain is simply
$2/3 \approx 0.666667$.
The naive Euclidean probability of separability is, then,
 $1/\sqrt{2} \approx 0.707107$.
Our attempts to compute with reasonable confidence
the corresponding Bures volumes
(and, thus, the Bures probability of separability) were again impeded by
numerical integration problems.

\section{Concluding Remarks}

Let us point out a number of papers \cite{ruth1,ruth2,ruth3}
(see also \cite[Chap. 6]{abstract}),
in which the volumes of {\it 
non-Euclidean} regular polytopes --- such as we 
have encountered in this study (secs.~\ref{sectA} \ref{sectE}) --- are 
computed.
For an (Euler angle) 
 parameterization of the $4 \times 4$ density matrices different from 
those
discussed above, see \cite{sudarmark}.
(This parameterization was used in the numerical integration
\cite{slatersilver}, in which the ``silver mean'' 
conjectures ((\ref{SDconjecture}), (\ref{silvermean})) 
were formulated, as the domain of possible values of the parameters is
readily expressible as a 15-dimensional hyperrectangle.)

Let us conclude on something of a discouraging note, 
however, in so far as we had been
hoping initially to be able to 
report some 
progress or conceptual breakthrough 
in {\it formally} resolving our
``silver mean conjectures'' ((\ref{SDconjecture}), (\ref{silvermean})), 
as to the statistical
distinguishability/Bures volumes of the 15-dimensional convex set of
separable two-qubit states.
(We continue to pursue the possibility of developing companion conjectures
for the 35-dimensional convex set of qubit-{\it qutrit} states \cite{qubitqutrit}, in terms of the
Bures and a number of other monotone metrics, as well as 
the Riemannian, but 
non-monotone Hilbert-Schmidt metric \cite{karolhans,ozawa}.)
The analyses presented above --- though perhaps of interest from a number
of perspectives --- appear to help little in this regard.

One possible further 
approach we have not yet investigated 
in any depth is the application of the
concept of {\it minimal} volume
\cite{bayard,bowditch,bambah}, seeing that the Bures metric does,  in fact,
serve as the {\it minimal} monotone metric \cite{petzsudar}.
(For a smooth manifold $M$, the minimal volume of $M$ is defined to be the greatest lower bound  of the total volumes of $M$ with respect to complete
Riemannian metrics, the {\it sectional} curvatures of which are bounded {\it 
above} in 
absolute value by 1 \cite{michael}. We note that 
{\it lower} bounds on the {\it scalar}
curvature are available from the work of Dittmann \cite{dittscalar}.)
In \cite{bayard}, demonstrating a conjecture of Gromov,
the minimal volume of $\bf{R}^2$ (the infinite Euclidean plane)
was shown
to be $\frac{2 \pi}{\sigma_{Ag}}$. (An exposition of this result is given
in \cite{bowditch}.) In \cite{bambah} the value of $\frac{1}{2 \sigma_{Ag}}$
 was obtained for a certain supremum of volumes.

\begin{acknowledgments}
I wish to express gratitude to the Kavli Institute for Theoretical
Physics for computational support in this research, and to A. Andai for
certain correspondence.

\end{acknowledgments}

\bibliography{Separable}

\end{document}